\documentclass[sigconf,nonacm]{acmart}
\AtBeginDocument{%
  }

\usepackage[normalem]{ulem}
\usepackage{float}
\usepackage{multirow}
\begin{document}

\title{PDMR: Passage-Driven Multi-ID Document Retrieval}

\author{Smail Oussaidene}
\email{smail.oussaidene@irit.fr}
\orcid{0000-0001-7747-0658}
\affiliation{%
  \institution{Institut de Recherche en Informatique de Toulouse (IRIT)}
  \city{Toulouse}
  \country{France}
  }

\author{Mohand Boughanem}
\email{mohand.boughanem@irit.fr}
\orcid{0000-0001-7004-0807}
\affiliation{%
  \institution{Institut de Recherche en Informatique de Toulouse (IRIT)}
  \city{Toulouse}
  \country{France}
  }

\renewcommand{\shortauthors}{S. Oussaidene et al.}

\begin{abstract}
Generative Retrieval (GR) models map queries directly to document identifiers, replacing conventional retrieval over external sparse or dense indexes with autoregressive identifier generation. However, most generative retrieval frameworks rely on a single-identifier assumption, mapping each document to a single target sequence. This forces the model to represent all document content with one sequence. Since documents are often multi-faceted, this can lead to lossy representations and reduced robustness to query variation, where multiple query intents must compete for a single generative access path.

In this work, we introduce Passage-Driven Multi-ID Retrieval (PDMR), a generative retrieval framework that represents documents through multiple passage-level identifiers. PDMR segments each document and assigns one identifier to each selected passage, which provides multiple semantic entry points for retrieving the same document. This multi-entry representation allows the model to align queries with specific semantic facets, thereby reducing the dependence on a single document-level target. To address the supervision ambiguity of this one-to-many mapping, we formulate training as a multi-target learning problem and explore an objective function designed to distribute probability mass across multiple valid passage-level identifiers.

We evaluate PDMR on NQ320K and MS MARCO Document. 
On NQ320K, PDMR improves over strong generative and non-generative baselines on Recall@1 and MRR@100. 
On MS MARCO Document, PDMR achieves the best Recall@1 and MRR@10 among the reported methods, while remaining competitive on Recall@10. 
Controlled ablations further show that passage-level supervision, identifier design, training-query augmentation, and multi-target learning contribute complementary gains.
\end{abstract}




\maketitle

\section{Introduction}

Generative Retrieval (GR) has emerged as a promising paradigm for information retrieval, replacing traditional index-based search with autoregressive generation of identifiers (DocIDs)
\cite{tayTransformerMemoryDifferentiable2022a, leeGLENGenerativeRetrieval2023a, zengScalableEffectiveGenerative2024, liMatchingGenerationSurvey2025}. 
Rather than retrieving documents via vector similarity or inverted index, \cite{karpukhinDensePassageRetrieval2020c, xiongApproximateNearestNeighbor2020c}, generative retrieval models produce document identifiers directly from the query \cite{tayTransformerMemoryDifferentiable2022a}.
In this setting, retrieval becomes a conditional text-generation task, shifting the system’s complexity from external storage components to the model itself.

A central design choice in generative retrieval is the construction of document identifiers.
Most existing approaches typically assign a {\bf single identifier} to each document, either as a numeric code or as a textual sequence derived from document content. 
Textual identifiers are constructed from surface elements of the document, including titles \cite{caoAutoregressiveEntityRetrieval2021}, n-grams \cite{bevilacquaAutoregressiveSearchEngines2022}, URLs \cite{zhouUltronUltimateRetriever2022}, or other salient units such as key terms \cite{leeGLENGenerativeRetrieval2023a}, thereby maintaining a direct correspondence with the original content.
In contrast, non-textual identifiers are generated through computational procedures such as vector quantization or hierarchical clustering \cite{wangNeuralCorpusIndexer2022, sunLearningTokenizeGenerative2023} to group semantically related documents.

The single-identifier assumption introduces a structural bottleneck, especially for long or multi-faceted documents. Compressing a document into a single generative target sequence leads to a lossy representation that fails to capture its diverse semantic aspects. Consequently, retrieval becomes sensitive to query variation: queries targeting specific facets may fail to retrieve the document if these aspects are not reflected in the global identifier, particularly under constrained decoding.

Recent research has sought to address this limitation, 
MINDER \cite{liMultiviewIdentifiersEnhanced2023a} 
introduces \textbf{multiview identifiers}, assigning different identifier types---such as titles, substrings, and pseudo-queries---to represent the same retrieval passage from various descriptive perspectives. 
Similarly, Few-Shot GR \cite{askariGenerativeRetrievalFewshot2025} introduces a one-to-many indexing strategy in which multiple docids are generated for the same document.
These studies show that multiple identifiers can improve generative retrieval by providing alternative access paths.  
However, such approaches operate at the level of a fixed retrieval unit (typically a document) and focus on enriching its representation. They do not address the structural challenge posed by document-level retrieval, where a single document may encompass multiple semantically distinct units.
Increasing the number of identifier views is not equivalent to increasing the number of semantic access points. While multiview approaches enrich how a unit is described, they do not change the granularity at which retrieval is performed. As a result, they do not resolve the core limitation of document-level generative retrieval: the inability to align diverse query intents with localized semantic content within a document.

To bridge this gap, we propose \textbf{Passage-Driven Multi-ID Retrieval (PDMR)}, a GR framework that represents each document using multiple identifiers corresponding to its internal semantic structure. 
Unlike ``view-driven'' methods that offer different descriptions of a whole, PDMR introduces multiple semantic units per document. This shift changes the retrieval target from one document identifier to several passage identifiers. As a result, the model enables retrieval through multiple semantic entry points.

To support this formulation, we cast training as a multi-target learning problem, where a query may be associated with multiple passage-level identifiers corresponding to relevant passages. 
We introduce a \textbf{weighted multi-target loss} specifically designed to distribute probability mass across multiple valid targets while prioritizing the most relevant passage.


Empirical results on NQ320K and MS MARCO Document show that PDMR improves over single-identifier baselines and remains competitive with strong retrieval methods.

Our contributions are summarized as follows:
\begin{itemize}
    \item  We introduce a passage-driven multi-identifier formulation of generative document retrieval, where each document is represented through identifiers assigned to selected internal passages rather than through a single global identifier.

    \item We propose a two-step LLM-based pipeline for passage extraction and the construction of \textbf{Title-Composed passage identifiers (TC-IDs)}.
    \item We formulate training as a multi-target learning problem, where multiple passage identifiers can serve as valid targets for the same relevant document.
    \item We provide experiments on NQ320K and MS MARCO Document, including ablations on passage-level supervision, identifier design, training-query augmentation, and multi-target learning.
\end{itemize}

\section{Related Work}
\subsection{Foundations of Generative Retrieval}

Generative retrieval formulates information retrieval as a sequence-to-sequence problem, where a model directly generates a symbolic identifier corresponding to a relevant document or passage given a query \cite{tayTransformerMemoryDifferentiable2022a, caoAutoregressiveEntityRetrieval2021}. 
Early work such as GENRE \cite{caoAutoregressiveEntityRetrieval2021} demonstrated that retrieval targets can be represented as natural-language entity names and generated autoregressively, while Differentiable Search Index (DSI) \cite{tayTransformerMemoryDifferentiable2022a} introduced the paradigm of treating a neural model itself as an index, trained to map queries directly to document identifiers.
Later work improves query-document alignment with synthetic query generation (DSI-QG), 
\cite{zhuangBridgingGapIndexing2023a}, 
learned discrete codes (NCI) \cite{wangNeuralCorpusIndexer2022}, 
 and n-gram identifiers (SEAL) \cite{bevilacquaAutoregressiveSearchEngines2022}). 
TOME \cite{renTOMETwostageApproach2023} uses tokenized URLs as identifiers and introduces a two-stage generation architecture to improve model-based retrieval. 
Other methods, such as GLEN \cite{leeGLENGenerativeRetrieval2023a}, GenRet \cite{sunLearningTokenizeGenerative2023}, and NOVO \cite{wangNOVOLearnableInterpretable2023}, further explore semantically meaningful identifiers, scalable training, and improved retrieval effectiveness.

\subsection{Identifier Design Strategies}

Document Identifier (DocID) design is a fundamental challenge in generative retrieval (GR), as the identifier must function both as a unique document address and a learnable target for sequence-to-sequence models.
Early research focused on static identifiers, which can be broadly
categorized into textual and non-textual forms. Textual designs utilize surface elements like
entity names or titles  \cite{caoAutoregressiveEntityRetrieval2021, tayTransformerMemoryDifferentiable2022a}, 
substrings \cite{bevilacquaAutoregressiveSearchEngines2022, liMultiviewIdentifiersEnhanced2023a}, 
pseudo-queries or synthetic queries \cite{zhuangBridgingGapIndexing2023a, liMultiviewIdentifiersEnhanced2023a}, 
lexical or keyphrase-like identifiers \cite{leeGLENGenerativeRetrieval2023a, wangNOVOLearnableInterpretable2023}, and LLM-generated free-text identifiers \cite{askariGenerativeRetrievalFewshot2025}. 
In contrast, non-textual designs rely on symbolic or learned identifiers, including atomic or numeric identifiers \cite{tayTransformerMemoryDifferentiable2022a}, semantically structured identifiers \cite{wangNeuralCorpusIndexer2022, zhouUltronUltimateRetriever2022}, and discrete learned document codes \cite{sunLearningTokenizeGenerative2023}.

A major research thread focuses on improving the semantic expressiveness of identifiers. GenRET \cite{sunLearningTokenizeGenerative2023} introduced a discrete auto-encoding approach to tokenize documents into short, learnable representations.
This direction has been further
refined by models like GLEN \cite{leeGLENGenerativeRetrieval2023a} and NOVO \cite{wangNOVOLearnableInterpretable2023} which optimize identifier learning to better capture document-specific semantics. 
More recently, MERGE \cite{zhangMultilevelRelevanceDocument2025} introduced a multi-level relevance identifier learning framework to   
learn high-quality discrete DocIDs that capture hierarchical semantic relationships between queries and documents.

Existing identifier designs trade off semantic interpretability and generation reliability: textual identifiers expose document content but can be long or ambiguous, while discrete or structured identifiers are easier to constrain but often provide weaker semantic signals \cite{tayTransformerMemoryDifferentiable2022a, caoAutoregressiveEntityRetrieval2021, wangNeuralCorpusIndexer2022, sunLearningTokenizeGenerative2023}. More importantly, most methods assume a single identifier per document, forcing multifaceted content into one sequence. This creates a structural bottleneck for long documents whose passages may cover different entities, events, or aspects, and makes retrieval brittle under autoregressive decoding: an early prefix error can move generation away from the correct identifier path, making the target document unreachable \cite{sunLearningTokenizeGenerative2023, zengScalableEffectiveGenerative2024, zhangGenerativeRetrievalTerm2024}. PDMR relaxes this one-to-one mapping through passage-driven decomposition.

\vspace{-0.2cm}
\subsection{Multiview versus Multi-Facet Identifiers}
To overcome the limitations of a single access point, recent efforts have explored the use of multiple identifiers. MINDER \cite{liMultiviewIdentifiersEnhanced2023a} introduced the concept of multiview identifiers, assigning different descriptive ``views' --such as titles, substrings, and pseudo-queries---to represent a single retrieval 
unit. 
Few-Shot GR \cite{askariGenerativeRetrievalFewshot2025} introduces few-shot indexing with a one-to-many mapping, where an LLM generates multiple free-text document identifiers without task-specific training. 
Although these approaches break the one-to-one mapping constraint, they mainly construct multiple views of the same unit rather than explicitly modeling the internal semantic structure of a document. 
These identifiers act as alternative descriptions of the same unit rather than as identifiers for distinct internal facets.
Another practical distinction is at inference time: MINDER queries the generator separately for each identifier view and then combines the generated title, substring, and pseudo-query evidence during ranking, whereas PDMR uses a single constrained decoding pass over one passage-identifier space.

PDMR differs by assigning identifiers to semantically distinct passages within each document, allowing document-level retrieval through localized access points.

\subsection{Fine-Grained Alignment Beyond Single Vectors}

Our motivation is also related to fine-grained representation learning in dense retrieval. 
Passage-level retrieval decomposes documents into smaller units, allowing queries to match localized evidence rather than a single global document representation \cite{karpukhinDensePassageRetrieval2020c}. 
Similarly, late-interaction models such as ColBERT \cite{khattabColBERTEfficientEffective2020b} and ColBERTv2 \cite{santhanamColBERTv2EffectiveEfficient2022} preserve token-level representations to support fine-grained query--document matching.
Our work adopts the same general principle in generative retrieval, but applies it to identifier design. 
Instead of relying on one global identifier per document, we derive multiple passage-driven identifiers that act as distinct semantic access points for document-level retrieval.

\section{Methodology}

\subsection{Task Formulation}
\label{sec:task_formulation}

Given a query $q$, generative retrieval (GR) aims to produce an identifier $y$ corresponding to a relevant document. 
This is modeled as an autoregressive generation process where the model predicts tokens one by one:

\[
p_\theta(y \mid q)
= \prod_{t=1}^{|y|}
p_\theta(y_t \mid q, y_{<t}),
\]
where $y = (y_1, \dots, y_{|y|})$ is a sequence of tokens representing the identifier.

In standard GR formulations, each document $d$ is associated with a single, unique identifier $y$. 
PDMR instead associates each document with a set of passage-level  identifiers:
\[
Z_d = \{y^{(1)}, \dots, y^{(K)}\},
\]
where each identifier, $y^{(i)}$ corresponds to a  specific semantic facet or passage within the document $d$ and $K$ denotes the total number of passages.

Let $d^*(q)$ denote a relevant document for query $q$, 
assuming that all passages of a relevant document are valid retrieval targets, the set of valid targets is therefore defined as:
\[
\mathcal{Y}(q) = Z_{d^*(q)} = \{y^{(1)}, \dots, y^{(K)}\}.
\]
Retrieval is performed by generating an identifier $\hat{y} \sim p_\theta(\cdot \mid q)$ and mapping it back to its corresponding document. 
This formulation allows any selected passage identifier to retrieve the document, creating multiple semantic entry points while keeping retrieval evaluated at the document level.

\subsection{Overview of PDMR}
\label{sec:pdmr_overview}

PDMR addresses two challenges: (1) the absence of explicit passage-level relevance labels, necessitating a mechanism to align queries with specific identifiers, and (2) the ambiguity of multiple valid targets, which requires a specialized training objective that can effectively handle such ambiguity.


PDMR addresses these challenges through four components:
(i) passage extraction to define fine-grained semantic units,
(ii) identifier construction to represent these units,
(iii) query–identifier alignment to bridge document-level supervision and passage-level targets,
and (iv) multi-target training to learn from multiple valid identifiers.

\subsection{Passage Extraction}
\label{sec:passage_extraction}

The first stage of the PDMR framework involves extracting a compact set of informative passages from each document. 
We first decompose each document into smaller, self-contained passages. 
This enables the model to associate different query intents with different parts of the document. 
We employ a two-step LLM-based procedure designed to balance coverage and precision.

\textbf{Initial Passage Extraction.}
Given a document $d$, an LLM is prompted to segment the document into a set of short passages. 
Each passage is represented by a title and a body, where the title summarizes the passage content and the body contains the passage text. 
Each passage is intended to capture a single idea, fact, or event. 
This first stage prioritizes recall: the resulting set may contain overlapping or low-importance passages, but it aims to cover all potentially relevant semantic units.

\textbf{Passage Selection and Refinement.}
In a second pass, the extracted passage set is provided back to the LLM, which is instructed to deduplicate and select only the most informative and distinct passages. By removing redundant or marginal segments, this step shifts the focus to precision, resulting in a final set of passages that forms the basis for identifier construction and fine-grained alignment

\subsection{Passage Identifier Construction}
\label{sec:passage_ids}

Following the semantic decomposition of documents into $K$ informative segments, PDMR assigns a distinct identifier $y^{(i)}$ to each extracted passage $p_i$.
Each document therefore receives multiple identifiers, each corresponding to a selected passage. 
If a document yields $K$ selected passages, it receives $K$ passage-level identifiers.

This design provides multiple semantic entry points, allowing the model to match specific query intents with localized passage facets. We investigate three identifier families to evaluate the impact of semantic structure on retrieval effectiveness.

\paragraph{(A) Random Numeric Passage IDs (RNIDs)}

The first identifier follows the "naive" identifier design introduced in the original DSI framework \cite{tayTransformerMemoryDifferentiable2022a}. 
Each passage is assigned a unique randomly generated numeric string that  is independent of its semantic content. 
These identifiers serve as purely symbolic targets for the generative model.
RNIDs serve as a symbolic baseline without semantic information in the identifier.

\paragraph{(B) Passage-Title Textual IDs (PT-IDs)}

The second identifier type assigns a textual identifier to each extracted passage.
For each passage, we reuse the title produced during passage extraction as its identifier.
While PT-IDs improve lexical alignment, they lack document-level context which may hinder the model to disambiguate between similar passages originating from different sources.

\paragraph{(C) Title-Composed Textual Passage IDs (TC-IDs)}
The third family constructs textual identifiers by combining the document title with the local  passage title.
This approach keeps both the document title and passage title in the identifier. We explore two composition strategies to determine how token ordering affects autoregressive decoding:

\begin{itemize}
    \item \textbf{Document-First Title ID (Doc$\rightarrow$Pass):} The identifier is formulated as: 
    \[ \text{document\_title} \,/\, \text{passage\_title}     \]
      This design provides a stable, shared prefix for all passages within the same document, allowing the model to first narrow the search space to the correct document before refining the prediction to a specific semantic unit\\

    \item \textbf{Passage-First Title ID (Pass$\rightarrow$Doc):} The identifier follows the format
    \[  \text{passage\_title} \,/\, \text{document\_title} \]
    prioritizing the localized semantic facet over the global context.
\end{itemize}


\subsection{Query--Identifier Alignment}
\label{sec:query_identifier_alignment}

A fundamental challenge in fine-grained generative retrieval is the discrepancy between the granularity of indexing targets and the granularity of available ground-truth labels.
While PDMR indexes documents through multiple passage-level identifiers to provide diverse semantic entry points, gold-standard relevance labels are typically provided only at the document level.
This creates a supervision misalignment: queries must be mapped to passage-level identifiers 

To create passage-level pairs, we construct our training set from two complementary sources: synthetic passage-based queries and original training queries aligned through document-level signals.

\begin{itemize}
    \item \textbf{Synthetic queries}: these queries are generated from passages using a query generation model (DocT5Query \cite{nogueiraDoc2queryDocTTTTTquerya}) and filtered to retain high-quality queries using a cross-encoder.
    
    Each query is generated from a specific passage rather than the full document. These queries serve as training inputs that map to the passage's unique ID, ensuring that every semantic unit has a dedicated generative access path. We refer to the source passage of a synthetic query as its \textit{primary passage} 
    
    \item \textbf{Training queries}: 
    Integrating original training queries is essential for exposing the model to natural information needs. However, since these queries are only associated with documents, their corresponding passage-level identifiers are not directly observed and must be inferred. We investigate two alignment strategies with different trade-offs:

    \begin{itemize}
    \item \textbf{(A) Document-to-All-Passages Assignment:} If a document is relevant to a query, we treat all of its internal passages as relevant targets. This maximizes supervision coverage and ensures that the model learns to associate the document’s global context with all its constituent facets, though it may introduce noise if a query only aligns with a specific subset of the document.
    \item \textbf{(B) Best-Matching Passage Assignment:} Alternatively, we select a single "best" passage per relevant document by computing the semantic similarity between the query and all passages using BM25 \cite{robertsonProbabilisticRelevanceFramework2009a}. This selective strategy provides a more precise and focused supervision signal, effectively reducing training-time noise at the cost of limited target coverage.
\end{itemize}

\end{itemize}







The final training set combines synthetic passage-generated queries with training queries constructed using one of these strategies.

\subsection{Multi-Target Training Losses}
\label{sec:multi_target_loss}
Representing documents through multiple passage-level identifiers introduces supervision ambiguity: a single query may semantically align with multiple segments within a relevant document.
While a query typically has a primary passage (i.e., the passage from which a synthetic query is generated, or the passage assigned to an original training query according to the chosen alignment strategy)
other facets of the same document may provide complementary evidence for achieving the final document-level retrieval objective.
At the same time, we aim to prioritize the passage that is the most directly aligned with the query. 
To address this, we replace the standard one-to-one generative mapping with a \textbf{multi-target learning problem}.

\paragraph{Pointwise Single-target formulation.}
As a baseline, we consider a standard single-target training setup, where each query is associated with a single identifier. 
In this setting, passage-level identifiers are treated as independent targets, and each query--identifier pair is considered as a separate training example. 
This corresponds to the conventional generative retrieval formulation, where the model is trained to predict a single identifier per query using a standard negative log-likelihood objective.


While simple to train, this approach does not treat the other identifiers of the same document as valid targets.

\paragraph{Multi-Target Formalism.}
For a query $q$ and its corresponding set of valid passage identifiers $\mathcal{Y}(q) = \{y^{(1)}, \dots, y^{(K)}\}$. 
Let $\ell_i$ be the negative log-likelihood (NLL) of target $y^{(i)}$ under the model:
\[
\ell_i = - \log p_\theta(y^{(i)} \mid q).
\]
A normalized weight vector $w = (w_1, \dots, w_K)$ specifies the relative contribution of each valid target, with $\sum_i w_i = 1$. 
This allows the training objective to emphasize the passage most aligned with the query while still accounting for other passages originating from the same document.
In contrast to the single-target setup, multi-target training associates each query with the identifier of its primary passage as well as additional identifiers corresponding to other passages from the same relevant document.
The multi-target loss is defined as the weighted average of the target losses:

\[
\mathcal{L}_{\text{avg}}
= \sum_{i=1}^{K} w_i \, \ell_i .
\]

This objective encourages the model to assign probability mass to multiple valid identifiers associated with the same document. 
Compared with the single-target formulation, it provides a more flexible supervision signal: the model is not trained only toward one passage identifier, but toward a weighted set of identifiers that all correspond to the same relevant document. 
At the same time, the weighting scheme preserves the distinction between the primary passage and less directly aligned passages.

\subsection{Inference and Document Ranking}
\label{sec:Inference}


During inference, PDMR uses a single constrained decoding pass
to generate candidate passage identifiers for a given test query $q$.
Unlike multiview approaches that run separate decoding procedures
for different identifier types, PDMR decodes over one unified
passage-identifier inventory. To ensure that the decoder only
produces valid targets belonging to this inventory, we employ
constrained beam search.

Because several passage identifiers can map to the same document, scores must be aggregated to produce a final document-level ranking. We define the relevance score of document $d$ as the maximum likelihood observed across its associated passage identifiers. Formally, for a document $d$ with a set of passage identifiers $Z_d$:
\[
\text{Score}(d, q) = \max_{y \in Z_d} p_\theta(y \mid q),
\]
where $p_\theta(y \mid q)$ is the joint probability of the generated token sequence for identifier $y$.

\section{Experimental Setup}

In this section, we introduce our experimental settings.

\subsection{Datasets}

We conduct experiments on two popular document retrieval benchmarks commonly used in generative retrieval: Natural Questions (NQ) and MS MARCO Document.

\paragraph{Natural Questions (NQ)}
For NQ, we use the NQ320K-style collection released with TSGen,\footnote{\url{https://github.com/namespace-Pt/TSGen}} which follows the standard setup used in recent generative retrieval studies. The collection is derived from Natural Questions and contains Wikipedia documents paired with natural language questions. After preprocessing, our corpus contains \textbf{109{,}739} documents, and the development split contains \textbf{7{,}830} queries. Following prior work on NQ320K, we also report results on the seen and unseen query splits, where the seen split contains queries whose target documents are associated with training queries, while the unseen split contains queries whose target documents have no labeled queries in the training data.

Applying the passage extraction procedure described in Section~\ref{sec:passage_extraction} yields a passage-level corpus of \textbf{428{,}412} passages. The resulting passages contain \textbf{67.15} words on average, with a median length of \textbf{66} words. Each document is represented by \textbf{3.90} passages on average, with a median of \textbf{4} passages per document.

\paragraph{MS MARCO Document (MSM Doc).}
We also evaluate on the MS MARCO Document collection, which consists of web documents associated with real search queries. Following prior generative retrieval work on MS MARCO Document, we construct a subset by retaining documents that are associated with at least one training or development query. This results in a corpus of \textbf{323{,}569} documents and \textbf{5{,}187} development queries.

We apply the same passage extraction procedure as for NQ, obtaining \textbf{1{,}142{,}073} passages. These passages contain \textbf{57.66} words on average, with a median length of \textbf{56} words. Each document is represented by \textbf{3.53} passages on average, with a median of \textbf{3} passages per document. Table~\ref{tab:passage_stats} summarizes the resulting passage-level corpora.

\begin{table}[t]
\centering
\caption{Statistics of the passage-level corpora constructed from NQ and MS MARCO Document.}
\label{tab:passage_stats}
\small
\begin{tabular}{lcccc}
\toprule
\textbf{Statistic} & \textbf{Min} & \textbf{Max} & \textbf{Mean} & \textbf{Median} \\
\midrule
\multicolumn{5}{l}{\textit{NQ}} \\
Passages per document & 1 & 16 & 3.90 & 4 \\
Passage length (words) & 23 & 287 & 67.15 & 66 \\
\midrule
\multicolumn{5}{l}{\textit{MS MARCO Document}} \\
Passages per document & 1 & 20 & 3.53 & 3 \\
Passage length (words) & 1 & 496 & 57.66 & 56 \\
\bottomrule
\end{tabular}

\end{table}

\subsection{Baselines}

We evaluate PDMR against three groups of baselines: sparse retrieval, dense retrieval, and generative retrieval methods. 
For NQ320K, we report results from the commonly reused NQ320K setting, following prior work. 
For MS MARCO Document, we use the baseline values reported in the comparison table of \cite{mekonnenLightweightDirectDocument2025b}.

\paragraph{Sparse retrieval.}
\textbf{BM25} \cite{robertsonProbabilisticRelevanceFramework2009a} is a lexical retrieval baseline based on exact term matching with term-frequency and inverse-document-frequency weighting. 
\textbf{DocT5Query} \cite{nogueiraDoc2queryDocTTTTTquerya} expands documents with synthetic queries generated by a sequence-to-sequence model, reducing vocabulary mismatch between queries and documents.

\paragraph{Dense retrieval.}
\textbf{DPR} \cite{karpukhinDensePassageRetrieval2020c} encodes queries and documents independently with dual encoders and retrieves documents by dense vector similarity. 
\textbf{ANCE} \cite{xiongApproximateNearestNeighbor2020c} improves dense retrieval through asynchronous hard negative mining, which provides stronger negative examples during training. 
\textbf{SentenceT5} \cite{niSentenceT5ScalableSentence2021} uses a T5-based encoder to produce dense text representations and serves as a strong neural retrieval baseline.

\paragraph{Generative retrieval.}
We compare against generative retrieval methods that differ mainly in how document identifiers are constructed.

\textit{Semantic and learned identifiers.}
\textbf{DSI} \cite{tayTransformerMemoryDifferentiable2022a} reformulates retrieval as sequence generation, where the model directly generates document identifiers. 
\textbf{DSI-QG} \cite{zhuangBridgingGapIndexing2023a} extends this framework with synthetic query generation to improve supervision coverage. 
\textbf{NCI} \cite{wangNeuralCorpusIndexer2022} uses hierarchical semantic identifiers and constrained decoding to improve scalability and retrieval accuracy.

\textit{Lexical and structured identifiers.}
\textbf{GENRE} \cite{caoAutoregressiveEntityRetrieval2021} performs retrieval through constrained generation of textual identifiers such as titles. 
\textbf{SEAL} \cite{bevilacquaAutoregressiveSearchEngines2022} uses n-grams from documents as identifiers, enabling retrieval through generated lexical spans. 
\textbf{TOME} \cite{renTOMETwostageApproach2023} generates structured identifiers such as URLs through a multi-stage framework. 
\textbf{Ultron} \cite{zhouUltronUltimateRetriever2022} explores document identifiers based on titles, URLs, and pseudo-queries.

\textit{Multi-view identifiers.}
\textbf{MINDER} \cite{liMultiviewIdentifiersEnhanced2023a} represents each retrieval unit with multiple identifier views, such as titles, substrings, and pseudo-queries. 
It is the closest baseline to PDMR, since both methods relax the single-identifier assumption. 
However, MINDER provides multiple views of the same retrieval unit, whereas PDMR assigns identifiers to different passage-level semantic units within a document.

We focus our experimental comparison on methods that are directly comparable to PDMR in terms of retrieval setting, benchmark usage, and identifier-based generative formulation. In particular, we compare against approaches that represent documents through generated identifiers, including semantic, lexical, hierarchical, and multi-view identifiers. Our goal is to isolate the effect of replacing a single document-level identifier with multiple passage-level identifiers under closely matched retrieval settings.


\subsection{Evaluation Metrics}

We report R@1, R@10, and MRR@100 for NQ320K, following prior generative retrieval work~\cite{leeGLENGenerativeRetrieval2023a}. 
For MS MARCO Doc, we report R@1, R@10, and MRR@10 \cite{mekonnenLightweightDirectDocument2025b}.

\subsection{Implementation Details}

All experiments are conducted using the \textbf{T5 Base} model as the backbone for generative retrieval. 
Models are trained for 3 epochs with a batch size of 256 on NVIDIA H100 80GB GPUs. 
We use the Adafactor optimizer with weight decay set to 0.01, gradient clipping with a maximum norm of 1.0, and label smoothing with a factor of 0.05. 
A linear warmup schedule is applied over 10\% of the total training steps. 
Training is performed in bfloat16 precision.
We tune the learning rate in the range $[3\mathrm{e}{-4}, 5\mathrm{e}{-4}]$ and report results using the best-performing configuration on the validation set.
For multi-target training, we employ a weighted loss over target identifiers. 
Given a query associated with multiple passages, we assign a weight of 0.6 to the primary, most relevant passage, while the remaining weight is uniformly distributed across the other associated passages. 
This gives higher weight to the primary passage while retaining the other identifiers as valid targets.
During evaluation, we use constrained decoding with beam search.
We use the \textbf{Qwen3-8B} model to segment documents into short passages.
For training-query augmentation, each original training query is replicated 20 times for each selected passage-level identifier.

For baselines, we follow common practice in generative retrieval research and compare PDMR against previously reported results from the literature using the same benchmark datasets, evaluation metrics, and underlying generation backbone whenever possible. 
In particular, our experiments follow the standard NQ320K and MS MARCO settings and employ the same T5 Base backbone used in prior work. 
While some implementation details and auxiliary components remain specific to each method, and exact reproduction of all baselines was not feasible, we carefully align the experimental setup to ensure comparison under the closest available setting.

\subsection{Dataset-level passage alignment}
\label{sec:dataset_alignment}

\begin{table}[!t]
\small
\caption{Passage-level alignment statistics for positive query-document pairs. Values are percentages.}
\label{tab:dataset_alignment}
\footnotesize
\setlength{\tabcolsep}{4pt}
\renewcommand{\arraystretch}{1.08}
\begin{tabular}{lcc}
\toprule
\textbf{Metric} & \textbf{NQ} & \textbf{MSM} \\
\midrule
Mean weak ratio @ 10\% & 10.05 & 24.20 \\
Mean weak ratio @ 25\% & 22.62 & 38.50 \\
Mean weak ratio @ 50\% & 44.76 & 52.42 \\
Pairs with weak ratio @ 25\% $\geq$ 50\% & 24.34 & 42.39 \\
\bottomrule
\end{tabular}
\end{table}

Before analyzing query-to-passage assignment, we examine how uniformly passages within a relevant document align with the query. 
For each positive query-document pair, we use BM25 to score all passages in the document against the query and compute the fraction of passages whose score is below a fixed percentage of the best passage score in that document. 
We call this the \textbf{weak passage ratio}; higher values indicate that the relevant document contains more passages that are weakly aligned with the query.

Table~\ref{tab:dataset_alignment} shows that relevant MSM documents contain more weakly aligned passages than relevant NQ documents. 
This difference is consistent with the source structure of the two datasets: NQ documents are Wikipedia articles, whereas MSM documents are full-length web pages and tend to contain more heterogeneous content. 
At the 25\% threshold, MSM has a mean weak passage ratio of 38.50\%, compared with 22.62\% for NQ. 
Moreover, 42.39\% of MSM positive pairs have at least half of their passages below this threshold, compared with 24.34\% for NQ. 
These statistics suggest that all-passage supervision may introduce more noisy targets on MSM than on NQ.

\subsection{Research Questions}
\label{sec:rqs}

Our evaluation focuses on four questions:

\begin{itemize}
    \item \textbf{RQ1:} Does passage-level supervision improve retrieval effectiveness compared to standard document-level identifiers?

    \item \textbf{RQ2:} How does the design of passage identifiers (random vs.\ textual, and composition strategies) affect retrieval performance?

    \item \textbf{RQ3:} What is the impact of training-query augmentation on generative retrieval, and how does it vary across datasets?

    \item \textbf{RQ4:} How does multi-target training objective influence model performance?
\end{itemize}
\section{Results and Analysis}
\label{sec:results}

We evaluate the proposed PDMR framework on document retrieval using standard ranking metrics, including Recall@1 (R@1), Recall@10 (R@10), and MRR@100.
We first present the overall comparison against strong baselines, and then analyze the contribution of each component through incremental controlled ablations, where components are added step by step to isolate their effect.

\subsection{Overall Results}
\label{sec:results_main}

\begin{table*}[t]
\centering
\caption[Results on NQ320K.]{
Results on NQ320K, including the full test set, seen queries, and unseen queries. We report R@1, R@10, and MRR@100, together with each method's backbone and identifier type. Abbreviations: Sem. = semantic, Hier. = hierarchical. 
Baseline numbers are collected from published papers under the commonly reused NQ320K setting. Results not available in the corresponding papers are denoted by ``--''.
Superscripts indicate the source from which the reported numbers are taken, not the original method paper: $^{G}$ from \cite{sunLearningTokenizeGenerative2023} and $^{T}$ from \cite{zhangGenerativeRetrievalTerm2024}.
}
\label{tab:nq320k_results}
\footnotesize
\setlength{\tabcolsep}{3pt}
\renewcommand{\arraystretch}{1.12}
\resizebox{\textwidth}{!}{
\begin{tabular}{ll l ccc ccc ccc}
\toprule
\textbf{Model} & \textbf{LLM} & \textbf{DocID}
& \multicolumn{3}{c}{\textbf{NQ320K}}
& \multicolumn{3}{c}{\textbf{Seen queries}}
& \multicolumn{3}{c}{\textbf{Unseen queries}} \\
\cmidrule(lr){4-6} \cmidrule(lr){7-9} \cmidrule(lr){10-12}
& & 
& \textbf{R@1} & \textbf{R@10} & \textbf{MRR@100}
& \textbf{R@1} & \textbf{R@10} & \textbf{MRR@100}
& \textbf{R@1} & \textbf{R@10} & \textbf{MRR@100} \\
\midrule

\multicolumn{12}{l}{\textit{Sparse \& dense retrieval}} \\

BM25 \cite{robertsonProbabilisticRelevanceFramework2009a}
& -- & Sparse
& 29.7 & 60.3 & 40.2
& 29.1 & 59.8 & 39.5
& 32.3 & 61.9 & 42.7 \\

DocT5Query \cite{nogueiraDoc2queryDocTTTTTquerya}
& T5 & Sparse
& 38.0 & 69.3 & 48.9
& 35.1 & 68.3 & 46.7
& 48.5 & 72.9 & 57.0 \\

DPR \cite{karpukhinDensePassageRetrieval2020c}
& BERT & Dense
& 50.2 & 77.7 & 59.9
& 50.2 & 78.7 & 60.2
& 50.0 & 74.2 & 58.8 \\

ANCE \cite{xiongApproximateNearestNeighbor2020c}
& RoBERTa & Dense
& 50.2 & 78.5 & 60.2
& 49.7 & 79.2 & 60.1
& 52.0 & 75.9 & 60.5 \\

SentenceT5 \cite{niSentenceT5ScalableSentence2021}
& T5 & Dense
& 53.6 & 83.0 & 64.1
& 53.4 & 83.9 & 63.8
& 56.5 & \textbf{79.5} & \underline{64.9} \\

\midrule
\multicolumn{12}{l}{\textit{Generative retrieval}} \\

GENRE$^{\mathrm{G}}$ \cite{caoAutoregressiveEntityRetrieval2021}
& BART/T5 & Titles
& 55.2 & 67.3 & 59.9
& 69.5 & 83.7 & 75.0
& 6.0 & 10.4 & 7.8 \\

DSI$^{\mathrm{G}}$ \cite{tayTransformerMemoryDifferentiable2022a}
& T5-base & Atomic/Sem.
& 55.2 & 67.4 & 59.6
& 69.7 & 83.6 & 74.7
& 1.3 & 7.2 & 3.5 \\

SEAL$^{\mathrm{G}}$ \cite{bevilacquaAutoregressiveSearchEngines2022}
& BART & n-gram
& 59.9 & 81.2 & 67.7
& -- & -- & --
& -- & -- & -- \\

MINDER$^{\mathrm{T}}$ \cite{liMultiviewIdentifiersEnhanced2023a}
& BART/T5 & Multi-view
& 62.7 & \textbf{86.9} & 71.3
& -- & -- & --
& -- & -- & -- \\

DSI-QG$^{\mathrm{G}}$ \cite{zhuangBridgingGapIndexing2023a}
& T5 & Sem.
& 63.1 & 80.7 & 69.5
& 68.0 & 85.0 & 74.3
& 45.9 & 65.8 & 52.8 \\

NCI$^{\mathrm{G}}$ \cite{wangNeuralCorpusIndexer2022}
& T5-base & Hier. sem.
& 66.4 & 85.7 & \underline{73.6}
& \underline{69.8} & \textbf{88.5} & \textbf{76.8}
& \underline{54.5} & 75.9 & 62.4 \\

TOME \cite{renTOMETwostageApproach2023}
& T5 & URL
& \underline{66.6} & -- & --
& -- & -- & --
& -- & -- & -- \\

\midrule
\textbf{PDMR (Ours)}
& T5-base & Passage multi-ID
& \textbf{67.6} & \underline{85.8} & \textbf{74.4}
& \textbf{70.1} & \underline{88.0} & \underline{76.7}
& \textbf{59.0} & \underline{78.2} & \textbf{66.6}
 \\

\bottomrule
\end{tabular}
}

\end{table*}

\begin{table}[t]
\centering
\caption[Results on MS MARCO Document.]{
Results on the MS MARCO document ranking dataset. We report Recall@1 (R@1), Recall@10 (R@10), and MRR@10, together with each method's backbone and identifier type. Baseline values are taken from the comparison table in \cite{mekonnenLightweightDirectDocument2025b}. 
}
\label{tab:msmarco_doc_results}
\footnotesize
\setlength{\tabcolsep}{4pt}
\renewcommand{\arraystretch}{1.15}
\begin{tabular}{ll l ccc}
\toprule
\textbf{Model} & \textbf{LLM} & \textbf{DocID} & \multicolumn{3}{c}{\textbf{MS MARCO Doc}} \\
\cmidrule(lr){4-6}
& & & \textbf{R@1} & \textbf{R@10} & \textbf{MRR@10} \\
\midrule

\multicolumn{6}{l}{\textit{Sparse \& dense retrieval}} \\

BM25 \cite{robertsonProbabilisticRelevanceFramework2009a}
& -- & Sparse & 18.94 & 55.07 & 29.24 \\

DocT5Query \cite{nogueiraDoc2queryDocTTTTTquerya}
& T5 & Sparse & 23.27 & 63.61 & 34.81 \\

DPR \cite{karpukhinDensePassageRetrieval2020c}
& BERT & Dense & 29.08 & 73.13 & 43.41 \\

ANCE \cite{xiongApproximateNearestNeighbor2020c}
& RoBERTa & Dense & 29.65 & 74.28 & 44.09 \\

Sentence-T5 \cite{niSentenceT5ScalableSentence2021}
& T5 & Dense & 27.27 & 72.15 & 40.69 \\

\midrule
\multicolumn{6}{l}{\textit{Generative retrieval}} \\

DSI \cite{tayTransformerMemoryDifferentiable2022a}
& T5-base & Sem. & 25.74 & 53.84 & 33.92 \\

DSI-QG \cite{zhuangBridgingGapIndexing2023a}
& T5 & Sem. & 28.82 & 62.26 & 38.45 \\

NCI \cite{wangNeuralCorpusIndexer2022}
& T5 & Hier. sem. & 29.54 & 67.28 & 40.46 \\

SEAL \cite{bevilacquaAutoregressiveSearchEngines2022}
& BART & n-gram & 27.58 & 61.01 & 37.68 \\

Ultron\cite{zhouUltronUltimateRetriever2022}
& T5 & Title, URL & 29.82 & 68.31 & 42.53 \\

Ultron \cite{zhouUltronUltimateRetriever2022}
& T5 & PQ & \underline{31.55} & 73.14 & \underline{45.35} \\

MINDER \cite{liMultiviewIdentifiersEnhanced2023a}
& T5 & Multi-view & 29.98 & \textbf{71.92} & 42.51 \\

\midrule
\textbf{PDMR (Ours)}
& T5-base & Passage multi-ID & \textbf{37.71} & \underline{71.49} & \textbf{48.66} \\

\bottomrule
\end{tabular}

\end{table}

Table~\ref{tab:nq320k_results} reports results on NQ320K, and Table~\ref{tab:msmarco_doc_results} reports results on MS MARCO Doc. 
Across both datasets, PDMR performs strongly among identifier-based generative retrieval methods.

On NQ320K, PDMR achieves 67.6 R@1, 85.8 R@10, and 74.4 MRR@100 on the full test set. 
It improves over DSI-QG by +4.5 R@1, +5.1 R@10, and +4.9 MRR@100, and over NCI by +1.2 R@1 and +0.8 MRR@100, while remaining close on R@10. 
These comparisons are relevant because DSI-QG and NCI also improve document identifiers, but still rely on a single access path per document, using semantic or hierarchical semantic identifiers. 
PDMR instead assigns multiple identifiers to different passages within the same document, which exposes several semantic entry points to the decoder.

The comparison with MINDER is particularly informative because it is the closest multi-identifier baseline. 
MINDER uses multiple views of the same retrieval unit, such as titles, substrings, and pseudo-queries, while PDMR assigns identifiers to different semantic units inside the document. 
PDMR improves over MINDER on R@1 and MRR@100, achieving 67.6 R@1 and 74.4 MRR@100 compared with 62.7 R@1 and 71.3 MRR@100. 
MINDER obtains a higher R@10, 86.9 compared with 85.8, suggesting that its multi-view design remains strong for broader top-$k$ recall. 
However, PDMR provides better early precision and ranking quality.

The seen/unseen split further supports this interpretation. 
On seen queries, PDMR obtains 70.1 R@1, 88.0 R@10, and 76.7 MRR@100, placing it among the strongest reported methods. 
On unseen queries, it reaches 59.0 R@1, 78.2 R@10, and 66.6 MRR@100, outperforming GENRE, DSI, DSI-QG, and NCI on all available unseen metrics. 
This suggests that passage-level identifiers help beyond memorizing observed query-document pairs, since they provide finer-grained access points for queries not seen during training.

On MS MARCO Doc, PDMR achieves 37.71 R@1, 71.49 R@10, and 48.66 MRR@10. 
It gives the best R@1 and MRR@10 among the reported methods, improving over strong generative baselines such as NCI, Ultron, and MINDER. 
Compared with Ultron, which uses title, URL, or pseudo-query identifiers, and MINDER, which combines multiple identifier views, PDMR obtains substantially higher first-rank accuracy. 
Its R@10 is slightly below the best reported value, but remains competitive with both dense and generative baselines. 
This mirrors the NQ320K trend: PDMR mainly improves the top of the ranking.

Overall, the results show that PDMR is competitive across both NQ320K and MSM Doc. 
Prior generative retrieval methods primarily improve the form of the document identifier, using titles, semantic identifiers, hierarchical codes, n-grams, pseudo-queries, URLs, or multiple views of the same document. 
PDMR instead changes the granularity of identifier assignment by moving from a single document-level target to multiple passage-level targets. 
The consistent gains in R@1 and MRR indicate that exposing multiple semantic access points improves early precision and ranking quality.

\subsection{RQ1: Passage-level supervision}
\label{sec:ablation_passage}

We first examine whether replacing document-level identifiers with passage-level identifiers improves retrieval effectiveness.
To isolate the effect of passage-level supervision, both variants in this ablation are trained with the same single-target loss; the only difference is whether the target identifier is defined at the document level or at the passage level.

\begin{table}[t]
\centering
\caption{Effect of passage-level supervision under single-target training. All values are percentages. 
$^{\dagger}$ indicates statistically significant improvement over Single-ID at $p<0.05$.}
\label{tab:ablation_passage}
\small
\setlength{\tabcolsep}{3.5pt}
\renewcommand{\arraystretch}{1.08}
\begin{tabular}{llcccccc}
\toprule
\textbf{Method} & \textbf{Target}
& \multicolumn{3}{c}{\textbf{NQ}}
& \multicolumn{3}{c}{\textbf{MSM}} \\
\cmidrule(lr){3-5} \cmidrule(lr){6-8}
&
& \textbf{R@1} & \textbf{R@10} & \textbf{MRR}
& \textbf{R@1} & \textbf{R@10} & \textbf{MRR} \\
\midrule
Single-ID & Doc ID
& 40.54 & 66.85 & 49.76
& 19.80 & 41.30 & 26.32
\\

PDMR & Pass. ID
& 45.62$^{\dagger}$ & 73.01$^{\dagger}$ & 55.36$^{\dagger}$
& 22.73$^{\dagger}$ & 52.27$^{\dagger}$ & 31.82$^{\dagger}$
\\
\bottomrule
\end{tabular}
\end{table}

Table~\ref{tab:ablation_passage} compares document-level and passage-level supervision on NQ and MSM under single-target training.
The Single-ID baseline assigns one identifier to each document, whereas PDMR assigns identifiers to passages and maps them back to the parent document at retrieval time.

Passage-level supervision improves all metrics on both datasets.
On NQ, PDMR increases R@1 from 40.54 to 45.62 (+5.08), R@10 from 66.85 to 73.01 (+6.16), and MRR from 49.76 to 55.36 (+5.60).
On MSM, the gains are also consistent: R@1 improves from 19.80 to 22.73 (+2.93), R@10 from 41.30 to 52.27 (+10.97), and MRR from 26.32 to 31.82 (+5.50).

These results support the passage-level formulation used in PDMR, even before introducing multi-target training.
A single document-level identifier forces the model to associate each document with one target sequence, which can be insufficient when documents contain multiple topics or answer-bearing regions.
Passage-level supervision provides multiple semantic access points for the same document, allowing queries to match more local content while still retrieving at the document level.

\subsection{RQ2: Identifier design}
\label{sec:ablation_ids}

We next examine how the design of passage identifiers affects retrieval performance.

\begin{table}[t]
\centering
\caption{Effect of identifier design. TC-ID denotes title-composed identifiers. All values are percentages. 
$^{\dagger}$ indicates statistically significant improvement over the corresponding RNID baseline at $p<0.05$.}
\label{tab:ablation_ids}
\scriptsize
\setlength{\tabcolsep}{3.5pt}
\renewcommand{\arraystretch}{1.08}
\begin{tabular}{llcccccc}
\toprule
\textbf{Method} & \textbf{Passage ID Type}
& \multicolumn{3}{c}{\textbf{NQ}}
& \multicolumn{3}{c}{\textbf{MSM}} \\
\cmidrule(lr){3-5} \cmidrule(lr){6-8}
&
& \textbf{R@1} & \textbf{R@10} & \textbf{MRR}
& \textbf{R@1} & \textbf{R@10} & \textbf{MRR} \\
\midrule
PDMR & RNID
& 45.62 & 73.01 & 55.36
& 22.73 & 52.27 & 31.82 \\

PDMR & PT-ID
& 48.81$^{\dagger}$ & 76.08$^{\dagger}$ & 58.58$^{\dagger}$
& 28.61$^{\dagger}$ & 59.11$^{\dagger}$ & 37.93$^{\dagger}$ \\

PDMR & TC-ID (Pass$\rightarrow$Doc)
& 49.34$^{\dagger}$ & 76.67$^{\dagger}$ & 59.05$^{\dagger}$
& 28.92$^{\dagger}$ & 60.00$^{\dagger}$ & 38.46$^{\dagger}$ \\

PDMR & TC-ID (Doc$\rightarrow$Pass)
& 53.60$^{\dagger}$ & 78.89$^{\dagger}$ & 62.89$^{\dagger}$
& 29.69$^{\dagger}$ & 59.86$^{\dagger}$ & 38.96$^{\dagger}$ \\

\midrule
PDMR & RNID + Multi-target (avg)
& 52.01 & 77.96 & 61.36
& 24.54 & 56.29 & 34.18 \\

PDMR & TC-ID (Doc$\rightarrow$Pass, avg)
& 54.60$^{\dagger}$ & 79.96$^{\dagger}$ & 63.91$^{\dagger}$
& 34.68$^{\dagger}$ & 66.05$^{\dagger}$ & 44.73$^{\dagger}$ \\

\bottomrule
\end{tabular}
\end{table}

Table~\ref{tab:ablation_ids} examines how passage identifier design affects retrieval performance on NQ and MSM. 
We compare random numeric identifiers (RNID), passage-title identifiers (PT-ID), and title-composed identifiers (TC-ID), with and without multi-target training.

Textual identifiers outperform RNID on both datasets. 
On NQ, PT-ID improves over RNID from 45.62 to 48.81 R@1, from 73.01 to 76.08 R@10, and from 55.36 to 58.58 MRR. 
On MSM, the gains are larger, with R@1 increasing from 22.73 to 28.61, R@10 from 52.27 to 59.11, and MRR from 31.82 to 37.93. 
This shows that replacing opaque numeric identifiers with passage-level textual identifiers provides useful semantic cues for query-passage matching.

Adding document context to passage identifiers improves performance, and the order of this context also matters. 
TC-ID (Pass$\rightarrow$Doc) improves over PT-ID on both datasets, reaching 49.34 R@1, 76.67 R@10, and 59.05 MRR on NQ, and 28.92 R@1, 60.00 R@10, and 38.46 MRR on MSM. 
Reversing the order to TC-ID (Doc$\rightarrow$Pass) gives the best single-target results, with 53.60 R@1, 78.89 R@10, and 62.89 MRR on NQ, and 29.69 R@1, 59.86 R@10, and 38.96 MRR on MSM. 
The gain is strongest on NQ, where Doc$\rightarrow$Pass improves over Pass$\rightarrow$Doc by +4.26 R@1, +2.22 R@10, and +3.84 MRR. 
On MSM, Doc$\rightarrow$Pass gives higher R@1 and MRR, while Pass$\rightarrow$Doc is slightly better on R@10. 
These results suggest that document-level context is useful in passage identifiers, and that placing it first often provides a stronger prefix for autoregressive decoding before the model selects the passage-level facet.

The bottom rows show that multi-target training is complementary to identifier design. 
For RNID, multi-target training improves NQ from 45.62 to 52.01 R@1 and MSM from 22.73 to 24.54 R@1. 
For TC-ID (Doc$\rightarrow$Pass), it gives the strongest results on both datasets, reaching 54.60 R@1, 79.96 R@10, and 63.91 MRR on NQ, and 34.68 R@1, 66.05 R@10, and 44.73 MRR on MSM. 

Overall, the results show that passage identifiers benefit from both semantic content and decoding-friendly structure. 
Textual identifiers improve query-identifier alignment, TC-ID adds useful document context, and the Doc$\rightarrow$Pass ordering provides a stronger generation prefix. 
Multi-target training further improves these identifiers by allowing several passage-level targets of the same document to contribute to learning.

\subsection{RQ3: Training-query augmentation}
\label{sec:ablation_queries}

We next investigate the impact of incorporating training queries as an additional supervision signal.
\begin{table}[t]
\centering
\caption{Effect of training-query augmentation. All values are percentages.
$^{\dagger}$ indicates statistically significant improvement over the corresponding Synthetic baseline at $p<0.05$.
$^{\ddagger}$ indicates statistically significant improvement over the other training-query assignment strategy at $p<0.05$.}
\label{tab:ablation_queries}
\scriptsize
\setlength{\tabcolsep}{3.2pt}
\renewcommand{\arraystretch}{1.08}
\begin{tabular}{llcccccc}
\toprule
\textbf{ID} & \textbf{Query aug.}
& \multicolumn{3}{c}{\textbf{NQ}}
& \multicolumn{3}{c}{\textbf{MSM}} \\
\cmidrule(lr){3-5} \cmidrule(lr){6-8}
&
& \textbf{R@1} & \textbf{R@10} & \textbf{MRR}
& \textbf{R@1} & \textbf{R@10} & \textbf{MRR} \\
\midrule
RNID & Synthetic
& 45.62 & 73.01 & 55.36
& 22.73 & 52.27 & 31.82 \\

RNID & Synthetic + Best-passage
& 60.13$^{\dagger}$ & 79.99$^{\dagger}$ & 67.57$^{\dagger}$
& 25.33$^{\dagger\ddagger}$ & 58.03$^{\dagger\ddagger}$ & 35.35$^{\dagger\ddagger}$ \\

RNID & Synthetic + All-passages
& 65.25$^{\dagger\ddagger}$ & 83.42$^{\dagger\ddagger}$ & 72.05$^{\dagger\ddagger}$
& 17.41 & 46.06 & 25.90 \\

\midrule
TC-ID & Synthetic
& 53.60 & 78.89 & 62.89
& 29.69 & 59.86 & 38.96 \\

TC-ID & Synthetic + Best-passage
& 66.17$^{\dagger}$ & 85.29$^{\dagger}$ & 73.48$^{\dagger}$
& 33.33$^{\dagger\ddagger}$ & 65.30$^{\dagger\ddagger}$ & 43.37$^{\dagger\ddagger}$ \\

TC-ID & Synthetic + All-passages
& 66.74$^{\dagger\ddagger}$ & 85.33$^{\dagger}$ & 74.32$^{\dagger\ddagger}$
& 27.24 & 57.55 & 36.59 \\
\bottomrule
\end{tabular}
\end{table}
Table~\ref{tab:ablation_queries} evaluates the effect of adding original training queries to synthetic passage-generated queries. 
We compare two alignment strategies: assigning each training query to its best-matching passage of the relevant document, and assigning it to all passages of that document.

On NQ, training-query augmentation brings large gains for both identifier types. 
For RNID, best-passage assignment improves R@1 from 45.62 to 60.13, R@10 from 73.01 to 79.99, and MRR from 55.36 to 67.57. 
Assigning queries to all passages performs even better, reaching 65.25 R@1, 83.42 R@10, and 72.05 MRR. 
The same trend holds for TC-ID: best-passage assignment reaches 66.17 R@1, 85.29 R@10, and 73.48 MRR, while all-passages assignment further improves to 66.74 R@1, 85.33 R@10, and 74.32 MRR. 
These results show that, on NQ, treating all passage identifiers of the relevant document as valid targets provides stronger supervision than selecting only one passage.

On MSM, best-passage training-query augmentation improves over the synthetic-only setting, while assigning each query to all passages degrades performance.
For RNID, best-passage assignment improves R@1 from 22.73 to 25.33, R@10 from 52.27 to 58.03, and MRR from 31.82 to 35.35. 
In contrast, all-passages assignment decreases performance to 17.41 R@1, 46.06 R@10, and 25.90 MRR. 
For TC-ID, best-passage assignment reaches 33.33 R@1, 65.30 R@10, and 43.37 MRR, while all-passages assignment reaches 27.24 R@1, 57.55 R@10, and 36.59 MRR.

This difference is consistent with the dataset-level passage alignment analysis in Table~\ref{tab:dataset_alignment}. 
NQ documents are Wikipedia articles, and their relevant documents contain fewer weakly aligned passages with respect to the query. 
In this setting, assigning a training query to all passages can expose the model to useful alternative passage identifiers of the same document. 
By contrast, MSM documents are full-length web pages and contain a larger fraction of weakly aligned passages inside relevant documents. 
Assigning a query to all passages on MSM can therefore introduce noisy targets, while best-passage assignment provides a more focused supervision signal.

Overall, training-query augmentation is an important component of PDMR, but the preferred assignment strategy depends on document structure. 
All-passages assignment is effective on NQ, where relevant documents are more uniformly aligned with the query. 
Best-passage assignment is preferable on MSM, where relevant documents often contain passages that are weakly related to the query.

\subsection{RQ4: Multi-target training objective}
\label{sec:ablation_multitarget}

We finally examine the impact of the proposed multi-target training objective.

\begin{table}[t]
\centering
\caption{Effect of multi-target training. TC-ID denotes the Doc$\rightarrow$Pass title-composed identifier. All values are percentages.
$^{\dagger}$ indicates statistically significant improvement over the corresponding Single-target baseline at $p<0.05$.
$^{\ddagger}$ indicates statistically significant improvement over the corresponding Multi-target avg baseline at $p<0.05$.}
\label{tab:ablation_multitarget}
\scriptsize
\setlength{\tabcolsep}{3.2pt}
\renewcommand{\arraystretch}{1.08}
\begin{tabular}{llcccccc}
\toprule
\textbf{ID Type} & \textbf{Objective}
& \multicolumn{3}{c}{\textbf{NQ}}
& \multicolumn{3}{c}{\textbf{MSM}} \\
\cmidrule(lr){3-5} \cmidrule(lr){6-8}
&
& \textbf{R@1} & \textbf{R@10} & \textbf{MRR}
& \textbf{R@1} & \textbf{R@10} & \textbf{MRR} \\
\midrule
RNID & Single-target
& 45.62 & 73.01 & 55.36
& 22.73 & 52.27 & 31.82 \\

RNID & Multi-target avg
& 52.01$^{\dagger}$ & 77.96$^{\dagger}$ & 61.36$^{\dagger}$
& 24.54$^{\dagger}$ & 56.29$^{\dagger}$ & 34.18$^{\dagger}$ \\

RNID & Multi-target avg + best-pass.
& 64.18$^{\dagger\ddagger}$ & 84.19$^{\dagger\ddagger}$ & 71.82$^{\dagger\ddagger}$
& 29.61$^{\dagger\ddagger}$ & 65.14$^{\dagger\ddagger}$ & 40.62$^{\dagger\ddagger}$ \\

RNID & Multi-target avg + all-pass.
& 65.57$^{\dagger\ddagger}$ & 84.11$^{\dagger\ddagger}$ & 72.64$^{\dagger\ddagger}$
& 20.24 & 54.00 & 30.37 \\

\midrule
TC-ID & Single-target
& 53.60 & 78.89 & 62.89
& 29.69 & 59.86 & 38.96 \\

TC-ID & Multi-target avg
& 54.60$^{\dagger}$ & 79.96$^{\dagger}$ & 63.91$^{\dagger}$
& 34.68$^{\dagger}$ & 66.05$^{\dagger}$ & 44.73$^{\dagger}$ \\

TC-ID & Multi-target avg + best-pass.
& 66.50$^{\dagger\ddagger}$ & 85.44$^{\dagger\ddagger}$ & 73.77$^{\dagger\ddagger}$
& 37.71$^{\dagger\ddagger}$ & 71.49$^{\dagger\ddagger}$ & 48.66$^{\dagger\ddagger}$ \\

TC-ID & Multi-target avg + all-pass.
& 67.59$^{\dagger\ddagger}$ & 85.81$^{\dagger\ddagger}$ & 74.45$^{\dagger\ddagger}$
& 30.02 & 62.21$^{\dagger}$ & 40.14$^{\dagger}$ \\
\bottomrule
\end{tabular}
\end{table}

Table~\ref{tab:ablation_multitarget} compares single-target training with the proposed multi-target average loss for both random numeric identifiers (RNID) and title-composed identifiers (TC-ID). 
The multi-target objective improves both identifier types on both datasets under synthetic-only supervision.

For RNID, multi-target training improves NQ from 45.62 to 52.01 R@1, from 73.01 to 77.96 R@10, and from 55.36 to 61.36 MRR. 
On MSM, it also improves performance, from 22.73 to 24.54 R@1, from 52.27 to 56.29 R@10, and from 31.82 to 34.18 MRR. 
Since RNID is purely symbolic, these gains show that the multi-target objective improves learning even when the identifiers do not contain semantic content.

For TC-ID, the synthetic-only gains are smaller on NQ but larger on MSM. 
On NQ, multi-target training improves R@1 from 53.60 to 54.60, R@10 from 78.89 to 79.96, and MRR from 62.89 to 63.91. 
On MSM, the gains are stronger, with R@1 increasing from 29.69 to 34.68, R@10 from 59.86 to 66.05, and MRR from 38.96 to 44.73. 
This indicates that multi-target learning complements semantic identifiers.

The strongest results are obtained when multi-target training is combined with training-query augmentation. 
On NQ, all-passages assignment gives the best performance for both RNID and TC-ID, reaching 65.57 R@1, 84.11 R@10, and 72.64 MRR for RNID, and 67.59 R@1, 85.81 R@10, and 74.45 MRR for TC-ID. 
On MSM, however, best-passage assignment is clearly stronger. 
With TC-ID, it reaches 37.71 R@1, 71.49 R@10, and 48.66 MRR, compared with 30.02 R@1, 62.21 R@10, and 40.14 MRR for all-passages assignment. 
The same pattern holds for RNID, where best-passage assignment outperforms all-passages assignment by +9.37 R@1, +11.14 R@10, and +10.25 MRR.
This contrast follows the dataset-level passage alignment analysis in Table~\ref{tab:dataset_alignment}.

Overall, multi-target improves both symbolic and textual passage identifiers, and its effect becomes strongest when combined with training-query augmentation. 
The best assignment strategy depends on dataset structure: all-passages assignment works best on NQ, while best-passage assignment is preferable on MSM.

\subsection{Summary of Findings}

The ablations show four main trends. 
First, passage-level supervision consistently improves over document-level supervision on both NQ and MSM. 
This confirms that representing a document with multiple passage-grounded identifiers provides more effective retrieval targets than using a single global document identifier.

Second, identifier design has a clear effect. 
Random numeric identifiers benefit from passage-level supervision, but textual identifiers perform better by exposing semantic cues. 
Among the single-target variants, TC-ID (Doc$\rightarrow$Pass) is the strongest overall, suggesting that placing document-level context before passage-level information provides a more effective decoding structure.

Third, training-query augmentation provides substantial gains, but the best assignment strategy depends on the dataset. 
On NQ, assigning each training query to all passage identifiers of the relevant document performs best, indicating that multiple passages often provide useful supervision for the same query-document pair. 
On MSM, best-passage assignment is stronger, consistent with the passage-alignment analysis showing that full-length web pages contain more weakly aligned passages. 
Thus, all-passages assignment can improve coverage when documents are more uniformly aligned with the query, but may introduce noisy targets when relevant documents are more heterogeneous.

Finally, multi-target training improves both RNID and TC-ID. 
The gains are visible under synthetic-only supervision and become larger when combined with training-query augmentation. 
By treating multiple passage identifiers of the same document as valid targets, the objective allows supervision to spread across several access paths while preserving the document-level retrieval goal.

Overall, these results support the design of PDMR. 
Documents are better retrieved through multiple passage-grounded identifiers than through a single document-level target, but the most effective query-to-passage assignment strategy depends on the internal alignment structure of the dataset.
\subsection{Limitations.}
PDMR improves over several identifier-based baselines, but it does not reach state-of-the-art performance on NQ320K. Recent methods such as GenRET \cite{sunLearningTokenizeGenerative2023}, NOVO \cite{wangNOVOLearnableInterpretable2023}, Few-Shot GR \cite{askariGenerativeRetrievalFewshot2025}, and TSGen \cite{zhangGenerativeRetrievalTerm2024} report stronger full-set results. However, these methods introduce factors outside the main focus of this work, such as learned identifier tokenization, optimized identifier representations, permutation-invariant term-set decoding, or larger backbone models. PDMR instead studies a controlled setting: replacing a single document-level identifier with multiple passage-level identifiers.

This focus also defines the limits of the approach. Passage-level identifiers provide multiple semantic access points and improve early precision and unseen-query generalization, but they do not address all challenges in generative retrieval. PDMR still relies on autoregressive decoding, which remains sensitive to early pruning and identifier ambiguity. It also uses fixed identifiers derived from document structure rather than identifiers learned end-to-end for retrieval. These results therefore position PDMR as complementary to identifier-learning and set-based decoding methods.

\section{Conclusion}
We introduced PDMR, a generative retrieval framework that represents each document with multiple passage-level identifiers. 
Instead of assigning a single identifier to an entire document, PDMR exposes several passage-based access points and maps them back to the document during retrieval. 
This design directly addresses the mismatch between multi-faceted documents and single-target generative retrieval.

Experiments on NQ320K and MS MARCO Doc show that passage-level identifiers improve retrieval effectiveness over document-level identifiers. 
The ablation studies further show that identifier design, training-query augmentation, and multi-target learning contribute complementary gains. 
Textual identifiers provide stronger query-identifier alignment than random identifiers, while the multi-target objective allows supervision to cover multiple valid passage identifiers for the same document. 
The results also show that query-to-passage assignment should account for dataset structure: all-passage assignment works best for NQ, whereas best-passage assignment is more effective for MSM, where relevant documents contain more weakly aligned passages.

Overall, our findings show that moving from single-entry to passage-grounded multi-entry document representations is an effective direction for generative retrieval. 
Future work can explore multi-stage training strategies that first learn reliable passage-level targets and then optimize document-level retrieval, as well as end-to-end frameworks that jointly learn passage selection, identifier construction, and query-identifier matching.

\begin{acks}
This work was performed using HPC resources from GENCI--IDRIS
(Grants 2025-105775 and 2025-103397).
\end{acks}

\appendix

\section{Training Computation}
\label{app:training_computation}

\begin{table}[t]
\centering
\caption{Training-set size and training time on NQ320K under different supervision settings. Relative size is computed with respect to the single-identifier training set.}
\label{tab:training_cost}
\footnotesize
\setlength{\tabcolsep}{4pt}
\renewcommand{\arraystretch}{1.08}
\begin{tabular}{lccc}
\toprule
\textbf{Setting} & \textbf{Instances} & \textbf{Relative size} & \textbf{Time} \\
\midrule
Single ID 
& 5.16M & 1.00$\times$ & 5.25 h \\

PDMR, single-target loss 
& 20.97M & 4.06$\times$ & 10.66 h \\

PDMR, multi-target loss 
& 20.97M & 4.06$\times$ & 41.53 h \\

PDMR + best-passage queries 
& 26.07M & 5.05$\times$ & 50.57 h \\

PDMR + all-passage queries 
& 45.70M & 8.86$\times$ & 91.53 h \\
\bottomrule
\end{tabular}
\end{table}

Table~\ref{tab:training_cost} reports the training-set size and training time on NQ320K under the main supervision settings. PDMR increases training computation through two sources. 
First, passage-level supervision expands the NQ320K training set because each document is represented by multiple passage identifiers rather than a single document identifier. 
In the single-target setting, this increases the training data from 5.16M to 20.97M instances, corresponding to a 4.06$\times$ increase over the single-ID baseline. 
When original training queries are added, the size further depends on the query-to-passage assignment strategy: best-passage assignment produces 26.07M instances, while all-passage assignment produces 45.70M instances.

Second, multi-target training increases the cost of each optimization step. 
Although the multi-target setting uses the same number of NQ320K instances as the single-target PDMR setting, its training time increases from 10.66 h to 41.53 h. 
This is expected because each query can be associated with multiple valid passage identifiers, and the loss must evaluate several targets for the same input.

Overall, on NQ320K, passage-level supervision mainly increases the number of training instances, while multi-target training increases the per-step computation.
\section{Illustration of Intra-Document Semantic Diversity}
\label{app:qualitative_example}

\begin{table}[t]
\centering
\caption{Example from MS MARCO Document where different queries for the same document align with different passage-level semantic units.}
\label{tab:qualitative_example_msm}
\footnotesize
\setlength{\tabcolsep}{2.5pt}
\renewcommand{\arraystretch}{1.05}
\begin{tabular}{p{0.25\columnwidth}p{0.36\columnwidth}p{0.30\columnwidth}}
\toprule
\textbf{Document} & \textbf{Passages} & \textbf{Attributed queries} \\
\midrule

\parbox[t]{0.25\columnwidth}{
\textbf{DocID:} \texttt{D83711}

\smallskip
\textit{Snippet:} The endocrine system consists of glands that secrete hormones directly into the bloodstream, including the pituitary, thyroid, parathyroid, adrenal, pancreas, gonads, and pineal glands...
}
&
\parbox[t]{0.36\columnwidth}{
\textbf{(1) \texttt{D83711\_5}: Pancreas}

\smallskip
Produces insulin and glucagon to regulate blood sugar levels and...

\vspace{0.45em}
\hrule
\vspace{0.45em}

\textbf{(2) \texttt{D83711\_7}: Pineal gland}

\smallskip
Located in the brain and secretes melatonin, regulating sleep-wake cycles...
}
&
\parbox[t]{0.30\columnwidth}{
\textbf{(1)} \textit{what gland secretes a substance that regulates glucose}

\vspace{1.45em}
\hrule
\vspace{0.45em}

\textbf{(2)} \textit{which gland is located closest to the human brain?}
}
\\

\bottomrule
\end{tabular}
\end{table}

Table~\ref{tab:qualitative_example_msm} shows an example from MS MARCO Document. 
Document \texttt{D83711} covers the endocrine system, but different queries align with different internal passages. 
This illustrates the motivation for PDMR: a single document-level identifier must compress these facets into one target sequence, while passage-level identifiers provide separate semantic access points.

\section*{GenAI Usage Disclosure}

ChatGPT (GPT-5) was used solely for language refinement, including checking spelling and improving grammar during the writing of this paper.

\bibliographystyle{ACM-Reference-Format}
\bibliography{references}

\end{document}